\newcommand{\ale}{\ \raisebox{-.3ex}{$\stackrel{<}{\scriptstyle \sim}$}\ }
\newcommand{\age}{\ \raisebox{-.3ex}{$\stackrel{>}{\scriptstyle \sim}$}\ }

\documentstyle{mn}

\input psfig

\title[Migration and the brown dwarf desert]
	{The brown dwarf desert as a consequence of \\ orbital migration}
	
\author[P.J. Armitage and I.A. Bonnell]{Philip J. Armitage and Ian A. Bonnell \\
	School of Physics and Astronomy, University 
	of St Andrews, North Haugh, St Andrews KY16 9SS}		
\begin{document}

\maketitle

\begin{abstract}
We show that the dearth of brown dwarfs in short-period orbits around 
Solar-mass stars -- the brown dwarf desert -- can be understood as a 
consequence of inward migration within an evolving protoplanetary disc.
Brown dwarf secondaries forming at the same time as the primary star 
have masses which are comparable to the initial mass of the 
protoplanetary disc. Subsequent disc evolution leads to inward migration, 
and destruction of the brown dwarf, via merger with the star. This is in 
contrast with massive planets, which 
avoid this fate by forming at a later epoch when the disc is close 
to being dispersed. Within this model, a brown dwarf desert arises 
because the mass at the hydrogen burning limit is coincidentally comparable 
to the initial disc mass for a Solar mass star. 
Brown dwarfs should be found in close binaries around very low mass stars, around 
other brown dwarfs, and around Solar-type stars during the earliest phases of star formation.
\end{abstract}

\begin{keywords}	
	accretion, accretion discs ---  planetary systems: protoplanetary 
	discs --- binaries: close --- stars: formation --- stars: low-mass, brown dwarfs	
\end{keywords}

\section{Introduction}
Recent surveys have demonstrated a high abundance of brown dwarfs in 
young clusters (Mart\'\i n et al. 2000), star forming regions (B\'ejar et al. 2001), 
and the field (Kirkpatrick et al. 1999, 2000). Brown dwarfs are also 
reasonably common in wide binaries (Gizis et al. 2001).
The sole well established exception is in close (semi-major axis $a \ale 4 \ {\rm AU}$) binaries.
The same radial velocity surveys that have been so successful in finding 
massive extrasolar planets show that brown dwarfs are rarely close binary companions to Solar-type 
stars (Marcy \& Butler 2000; Halbwachs et al. 2000). This brown dwarf desert 
supports the conventional belief 
that massive planets and brown dwarfs form in distinctly different 
ways, and also suggests that some aspect of the 
formation or early evolution of brown dwarfs differs from that of 
stars. 

One possibility for explaining the desert is to postulate 
that the formation mechanism for brown dwarfs is entirely different 
from that of {\em either} stars or giant planets. Reipurth \& Clarke (2001), 
for example, suggest that brown dwarfs are objects whose growth 
towards stellar masses was curtailed by ejection from small 
multiple systems (see also Reipurth, Clarke \& Delgado-Donate 2001). 
In their model, the desert arises because substellar objects 
that were {\em not} ejected continued to accrete, eventually reaching stellar masses.

In this paper, we investigate a less radical (and more limited) possibility. 
We show that the absence of brown dwarfs in close binaries with Solar-type stars 
could be due to orbital migration within an evolving protoplanetary disc. This is the same 
process that is invoked to explain the presence of massive planets 
at small orbital radii (Lin, Bodenheimer \& Richardson 1996), and which may also 
play a role in the orbital evolution of cataclysmic variables  
(Taam \& Spruit 2001) and supermassive black hole binaries (Gould \& Rix 2000). The 
critical assumption is that brown dwarfs form contemporaneously with 
the star, and thus become embedded within a young and relatively massive 
protoplanetary disc. Under these conditions, 
we show that migration efficiently clears out the desert by forcing the brown 
dwarfs into mergers with the star. This is in contrast to the evolution 
of massive planets, which have a chance of escaping the same fate by forming 
later, when the disc is close to being dispersed and cannot drive migration 
through to merger (Trilling, Lunine \& Benz 2001; Armitage et al. 2001).

\section{Brown dwarf migration}

We assume that close brown dwarfs companions form at essentially the same time as the 
star, either via fragmentation during cloud collapse, or via disc instabilities 
during the earliest protostellar stages (e.g. Bonnell \& Bate 1994). The brown dwarf 
companions open up a gap in the protoplanetary disc, and their subsequent fate 
then depends upon the evolution of the coupled disc-brown dwarf system. If the 
disc mass is large enough, relative to the brown dwarf mass, then the brown 
dwarf migrates in radius as the disc evolves as if it were a fluid element 
in the gas. This can be inward or outward, depending upon the sign of the 
radial velocity in the disc. If the disc mass is too low, on the other 
hand, the brown dwarf will act as a dam in the disc, holding up accretion 
and suffering minimal orbital migration (Syer \& Clarke 1995). We show 
later that the first regime, where migration is rapid, is probably 
appropriate for brown dwarfs around Solar mass stars, while the no-migration 
outcome is likely if the star has very low mass.

\subsection{Numerical methods}

The interaction of a low mass secondary with particulate or gaseous discs 
has been extensively studied (Goldreich \& Tremaine 1980; Lin \& Papaloizou 1986).
Based on this work, we model the interaction using a one dimensional treatment, in 
which we solve for the coupled evolution of the disk surface density 
$\Sigma (r,t)$ and brown dwarf semi-major axis $a$. The governing 
equation, including mass loss in a disc wind at a rate 
$\dot{\Sigma}_w(r)$, is (Lin \& Papaloizou 1986; Trilling et al. 1998),
\begin{equation}
 { {\partial \Sigma} \over {\partial t} } =
 { 1 \over r } { \partial \over {\partial r} }
 \left[ 3 r^{1/2} { \partial \over {\partial r} }
 \left( \nu \Sigma r^{1/2} \right) -
 { { 2 \Lambda \Sigma r^{3/2} } \over
 { (G M_*)^{1/2} } } \right] - \dot{\Sigma}_w.
\label{eqsigma}
\end{equation}  
where $M_*$ is the stellar mass. 
The first term on the right-hand side describes the diffusive 
evolution of a disk in which the angular momentum transport can 
be parameterized via a kinematic viscosity $\nu$. The second 
term describes how the disk responds to the torque from the 
secondary, which is approximated as a fixed function of radius 
$\Lambda (r,a)$. We adopt,
\begin{eqnarray}
 \Lambda & = & - { {q^2 G M_*} \over {2 r} }
 \left( {r \over \Delta_p} \right)^4 \, \, \, \, r < a \nonumber \\
 \Lambda & = & { {q^2 G M_*} \over {2 r} }
 \left( {a \over \Delta_p} \right)^4 \, \, \, \, r > a
\label{eqtorque}
\end{eqnarray}
where $q = M_{\rm BD} / M_*$ 
is the mass ratio of the binary, and $\Delta_p$ is 
given in terms of the disc scale height $h$ by,
\begin{equation}
 \Delta_p = {\rm max} ( h, \vert r - a \vert ).
\end{equation}    
The rate of migration of the brown dwarf is,
\begin{equation}
 { { {\rm d} a } \over { {\rm d} t } } =
 - \left( { a \over {GM_*} } \right)^{1/2}
 \left( { {4 \pi} \over M_{\rm BD} } \right)
 \int_{r_{\rm in}}^{r_{\rm out}}
 r \Lambda \Sigma {\rm d} r.
\end{equation}  
Tests show that the detailed form of the torque function is unimportant 
for determining the migration rate once a gap has been 
opened. We have used the same form adopted by 
Armitage et al. (2001) to study massive planet migration.

The use of equation (\ref{eqsigma}) implicitly assumes that 
(a) the brown dwarf orbit remains circular, and (b) that 
there is no ongoing accretion across the gap. Current 
theoretical understanding does not permit a definitive 
answer to either question. Existing simulations 
suggest that the assumption of circular orbits ought 
to be reasonable for massive planets, and perhaps for 
low mass brown dwarfs, but that for more massive brown 
dwarfs ($M_{\rm BD} \age 20 M_{\rm Jupiter}$) or stars the interaction 
with the disc may drive eccentricity growth (Artymowicz et al. 1991; 
Papaloizou, Nelson \& Masset 2001). Ongoing accretion is likely 
to be substantially reduced at the relatively large 
mass ratios appropriate to brown dwarf secondaries 
(Bryden et al. 1999), but the exact level is rather 
uncertain.

\subsection{Disc model}

For the disc, we use a variant of the Clarke, Gendrin \& 
Sotomayor (2001) model, which combines viscous evolution with
mass loss at large radii. The model is motivated by observations 
of disc photoevaporation in Orion (Johnstone, Hollenbach \& 
Bally 1998), though the same process may also occur at a lower 
level for relatively isolated stars (Shu, Johnstone \& Hollenbach 1993). 
Including mass loss from the disc enables 
the model to reproduce the observed rapid transition 
between accreting Classical T Tauri stars and non-accreting 
Weak-Lined systems (Clarke, Gendrin \& Sotomayor 2001). We take,
\begin{equation}
 \nu = 1.75 \times 10^{13} \left( { r \over {1 \ {\rm AU}} } \right)^{3/2} \ {\rm cm^2 s^{-1}}
\end{equation} 
consistent with a $\Sigma \propto r^{-3/2}$ surface density profile.
We start the runs with a steady-state surface density profile 
(accretion rate constant with radius), and assume 
$h = 0.05 r$ at all radii. 

Appropriate values for the mass loss are poorly known for the 
situation, relevant here, where the star is {\em not} part of 
a rich cluster such as Orion. We assume that 
mass is lost from the disc outside a critical radius 
$r_{\rm crit} = 5 \ {\rm AU}$, with a radial scaling 
$\dot{\Sigma}_w \propto r^{-1}$. For most of our runs, 
we normalise $\dot{\Sigma}_w$ such 
that the total mass loss rate integrated to 25~AU is 
$5 \times 10^{-9} \ M_\odot {\rm yr}^{-1}$. To gauge the 
sensitivity of the results to the mass loss prescription, we also 
run models with the mass loss rate reduced to 
$5 \times 10^{-10} \ M_\odot {\rm yr}^{-1}$.

\begin{figure}
\psfig{figure=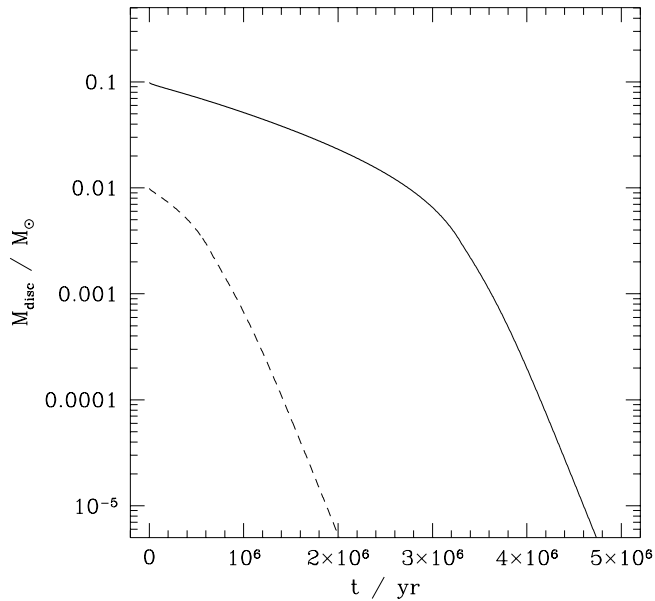,width=3.5truein,height=3.5truein}
\caption{Evolution of the disc mass with time for the two disc 
	models, which differ only in the assumed initial conditions. 
	The solid line shows a model with an initial mass $M_{\rm init} = 0.1 \ M_\odot$, 
	and an accretion rate $\dot{M}_{\rm init} = 3.3 \times 10^{-8} \ M_\odot$.
	The dashed line is for a model with $M_{\rm init} = 10^{-2} \ M_\odot$, 
	and $\dot{M}_{\rm init} = 3.3 \times 10^{-9} \ M_\odot$. The disc 
	is dispersed rapidly once the disc wind becomes significant.}
\label{fig1}	
\end{figure}

Figure 1 shows the evolution of the disc mass for initial masses 
of $0.1 \ M_\odot$ and $10^{-2} \ M_\odot$. We concentrate on the 
higher initial disc mass case, which is likely to be representative of the 
disc around a Solar mass star. A disc mass of the order of $\sim 0.1 \ M_*$ is
comparable to the most massive discs inferred from mm wavelength 
observations (Osterloh \& Beckwith 1995), and can also be justified theoretically -- it
corresponds to the marginally gravitationally unstable state 
expected as an endpoint of the disc formation process (e.g. Lin \& 
Pringle 1990). For this model, a break in the 
evolution occurs at around 3~Myr, when the disc wind becomes dominant. The subsequent 
evolution is consistent both with the inferred time-dependence of the accretion  
rate for T Tauri stars (Hartmann et al. 1998), and with the 
evolution of the disc fraction with cluster age (Haisch, Lada \& Lada 2001). 

Operationally, we solve equation (\ref{eqsigma}) using an explicit 
method on a non-uniform grid. The runs described below all used 
300 mesh points, with an inner radius of 0.075~AU and an outer 
radius of 33.3~AU. At $R_{\rm in}$ a zero-torque boundary 
condition ($\Sigma = 0$) was used, while at $R_{\rm out}$ 
we set the radial velocity $v_r = 0$. 

In our current implementation, brown dwarfs reaching 
the inner boundary at $R_{\rm in}$ are declared as mergers. 
We do not include any `stopping mechanism', sometimes invoked 
to slow or halt migration of extrasolar planets at very small radii. 
This is because the evidence for such mechanisms is weak. 
Although there are plausible reasons why migration might 
stall at small radii -- for example due to an inner 
magnetospheric cavity in the disc (K\"onigl 1991) -- 
the observed radial distribution of extrasolar planets 
can be matched satisfactorily {\em without} including 
any stopping mechanism (Trilling et al. 2001; Armitage et al. 
2001).

\subsection{Migration-driven mergers}

\begin{figure}
\psfig{figure=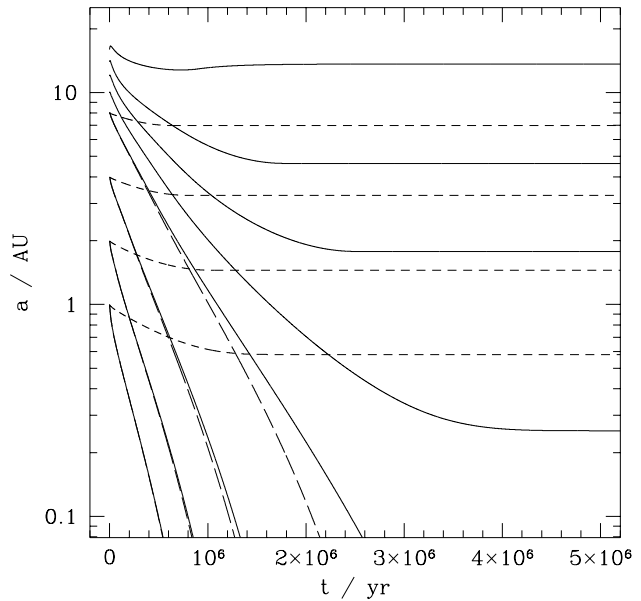,width=3.5truein,height=3.5truein}
\caption{Orbital migration of a $0.04 \ M_\odot$ brown dwarf in a 
	protoplanetary disc. The solid curves show how brown dwarfs with 
	different initial orbital radii migrate in discs 
	with an initial mass of $0.1 \ M_\odot$. All brown dwarfs 
	with initial orbital radii $a < 10 \ {\rm AU}$ are swept 
	inside 0.1~AU, and probably lost in mergers. We also show 
	how some of the results depend on the disc model. Reducing the mass 
	loss rate from the disc (long dashed curves) makes little 
	difference to the migration history, while reducing the 
	initial disc mass to $10^{-2} \ M_\odot$ (short dashed curves) 
	almost halts inward migration.}
\label{fig2}	
\end{figure}

Figure 2 shows the migration history of brown dwarfs embedded in 
protoplanetary discs with initial masses of $0.1 \ M_\odot$ and $10^{-2} \ M_\odot$. 
The brown dwarf mass was $0.04 \ M_\odot$, 
i.e. substantially less than the disc for the more massive initial 
disc model. Disc evolution in the initially more massive disc drives the brown 
dwarf into merger with the star for all initial orbital radii $a \ale 10 \ {\rm AU}$.
Migration is especially rapid for brown dwarfs with initial orbits 
inside $\sim 2 \ {\rm AU}$. These objects are swept inside $0.1 \ {\rm AU}$  
within less than a Myr after the start of the calculation. Reducing the 
rate of mass loss from the outer disc by an order of magnitude 
tends to {\em increase} the extent 
of migration, as it means that there is a larger reservoir of gas exterior to the 
brown dwarf orbit at late times. For orbits with initial semi-major 
axis $a \ale 10 \ {\rm AU}$, however, the change is relatively small.

As expected, this rapid migration does not occur if the disc mass 
is much smaller than the brown dwarf mass. For a disc with an initial 
mass of only $10^{-2} \ M_\odot$, the angular momentum reservoir 
of the brown dwarf is large enough to prevent substantial orbital 
evolution. Only a modest migration occurs before the disc wind 
disperses the disc and freezes the brown dwarf into its final orbit. 
The low mass runs provide a qualitative idea of the evolution around 
very low mass stars, where the initial disc mass is likely to be smaller 
than the mass of any brown dwarfs present. If we assume that the initial 
disc mass is proportional to the stellar mass, then the results suggest 
that substantial migration would still occur for moderately less 
massive stars than the Sun, with masses 
$\approx 0.5 \ M_\odot$. For stars with masses of $0.1 - 0.2 \ M_\odot$,  
however, very little migration is expected.

Migration will efficiently empty the desert of brown dwarfs with initial 
semi-major axis $a \ale 8 \ {\rm AU}$, provided that the initial disc mass 
is of the order of $0.1 \ M_\odot$. The only brown dwarfs that will 
remain at radii observable with radial velocity surveys are those
that have migrated inwards from greater distances, before becoming stranded 
at small separations by the dispersal of the disc. Continuity demands that 
such objects must exist, but their numbers are 
generically likely to be small. In most models the disc gas is outflowing 
at large radii (Basu 1998; Hartmann et al. 1998) and tending to drive 
outward, rather than inward, migration. Moreover, as is evident from the 
divergence of the tracks in Fig.~2, migration leads a dilution in the 
number of brown dwarfs per unit logarithm in radius. Surveys have discovered some 
brown dwarfs at small radii -- for example the outer companion (at $\approx 3 \ {\rm AU}$) 
of the HD~168443 system (Marcy et al. 2001) -- so the observations require 
only a substantial rather than a complete depletion of the population.
Quantitatively, Gizis et al. (2001) estimate that if brown dwarfs 
followed the same distribution of separations as stars, at least ten 
times the observed number would be expected with 
separations less than 3~AU. There is considerable uncertainty in this 
estimate, but it nonetheless suggests that a theory for the desert needs 
to produce roughly an order of magnitude depletion in the population of short period 
brown dwarfs. 

To estimate the range of initial orbital radii that will lead to surviving 
short-period brown dwarfs, we have run a series of models in which brown 
dwarfs were placed at larger initial radii in a $0.1 \ M_\odot$ disc. 
The results of these calculations are also shown in Fig.~2. For our disc model, 
binaries with initial orbital radii in the range 
$10 \ {\rm AU} \ale a \ale 14 \ {\rm AU}$ are progenitors of observable 
systems in which the final orbital radius is $a < 4 \ {\rm AU}$. Smaller 
initial separations lead to merger, as discussed already, while larger 
separations fail to migrate inwards into the observable window prior 
to disc dispersal.

To use these results to estimate the net depletion of the desert, we 
need to know the initial distribution of brown dwarf orbital radii. 
If we assume that the formation of binaries that include a brown 
dwarf is scale free (i.e. equal numbers per logarithmic interval of 
separation), then migration leads to a net depletion by an order of 
magnitude in the $0.1 \ {\rm AU} < a < 4 \ {\rm AU}$ range of orbital 
radii. Alternatively, we could assume that the initial radii of 
the binaries mirrored that observed for solar-type main sequence 
binaries generally. Duquennoy \& Mayor (1991) parameterize the 
distribution with period $P$ (in days) as,
\begin{equation}
 f(\log P) \propto e^{- \left[ {(\log P - \overline{\log P})^2} \over 
 {2 \sigma^2_{\log P} } \right] },
\end{equation}    
where $\overline{\log P} = 4.8$ and $\sigma_{\log P} = 2.3$. This 
distribution is an increasing function over the range of initial 
radii of interest here, and thus we estimate a rather smaller depletion 
of brown dwarfs using this as the initial distribution. Between 
$0.1 \ {\rm AU} < a < 4 \ {\rm AU}$, we obtain a depletion by 
a factor of 6.

\subsection{Rotation rates of merger remnants}

The best observational test of this model would be the detection 
of a significantly larger brown dwarf frequency among close 
pre-main-sequence binaries, in which accretion is still ongoing. 
There may also be evidence for mergers within the distribution 
of stellar rotation rates. Unlike gas, which can be accreted 
via a magnetosphere with essentially zero net change in stellar 
specific angular momentum (K\"onigl 1991; Armitage \& Clarke 1996), 
a merging brown dwarf arrives with the Keplerian angular 
momentum near the stellar surface. The ratio of the brown dwarf 
to stellar angular momentum is,
\begin{equation}
 { L_{\rm BD} \over L_* } = {1 \over k^2} \left( {M_{\rm BD} \over M_*} \right) 
 \left( {\Omega_K \over \Omega_*} \right)
\end{equation}
where $\Omega_*$ is the stellar angular velocity, $\Omega_K$ the 
Keplerian angular velocity at the stellar surface, and $k^2 \approx 0.2$ 
is the radius of gyration for the star (assumed fully convective). 
For an initial rotation period $P \sim 7 \ {\rm dy}$, appropriate 
to many Classical T Tauri stars (Bouvier et al. 1993; but see also 
Stassun et al. 1999), $ L_{\rm BD} / L_* \gg 1$, so the 
merger leads to substantial spin-up. The results outlined above 
suggest that of the order of 10\% of stars might have suffered 
such mergers with brown dwarfs. If the spin-down time-scale 
subsequent to merger is relatively long (of the order of a Myr), 
we would then expect a small fraction (a few percent) of Classical 
T Tauri stars to be rapid rotators as a consequence of mergers.

\section{Summary}

We have shown that inward orbital migration of brown dwarfs, 
within an evolving protoplanetary disc, can account for a low 
frequency of brown dwarfs as close binary companions to stars 
with masses $M_* \sim M_\odot$. Migration depletes the initial 
frequency of brown dwarf companions at all radii where there  
is significant viscous evolution of the protoplanetary disc. 
This region could extend out to several tens of AU. Brown dwarfs with 
initially larger orbital radii $a \sim 10^2 \ {\rm AU}$ would 
be unaffected. In our specific model, migration leads to the destruction 
of all brown dwarfs with initial orbital radii $a \ale 10 \ {\rm AU}$,
via mergers with the star which cause significant stellar spin-up. 
Orbits with $0.1 \ {\rm AU} < a < 4 \ {\rm AU}$ are partially 
replenished by brown dwarfs migrating inwards from still greater 
radii, but the net brown dwarf frequency at these radii is still reduced by 
a factor of 5-10.

The main prediction of the model is that brown dwarfs in close 
orbits ought to be up to an order of magnitude more common 
amongst the youngest pre-main-sequence stars (less than a Myr), 
as compared to the main sequence. For migration to occur, we 
also require that the initial disc mass be at least 
comparable to the mass of the brown dwarf. No brown dwarf desert 
is thus expected around the lowest mass hydrogen-burning stars 
($0.1 - 0.2 \ M_\odot$), or indeed around other brown dwarfs, whose 
discs would have been too feeble to drive significant brown dwarf 
migration.


\begin{thebibliography}{}

\bibitem{}
 Armitage P.J., Clarke C.J., 1996, MNRAS, 280, 458

\bibitem{}
 Armitage P.J., Livio M., Lubow S.H., Pringle J.E., 2001, MNRAS, submitted
 
\bibitem{}
 Artymowicz P., Clarke C.J., Lubow S.H., Pringle J.E., 1991, ApJ, 370, L35
 
\bibitem{}
 Basu S., 1998, ApJ, 509, 229 

\bibitem{}
 B\'ejar V.J.S. et al., 2001, ApJ, 556, 830
 
\bibitem{}
 Bonnell I.A., Bate M.R., 1994, MNRAS, 271, 999
 
\bibitem{}
 Bouvier J., Cabrit S., Fernandez M., Martin E.L., Matthews J.M., 
  1993, A\&A, 272, 176
  
\bibitem{}
 Bryden G., Chen X., Lin D.N.C., Nelson R.P., Papaloizou J.C.B., 1999, 
 ApJ, 514, 344  
 
\bibitem{}
 Clarke C.J., Gendrin A., Sotomayor M., 2001, MNRAS, 328, 485

\bibitem{} 
 Duquennoy A., Mayor M., 1991, A\&A, 248, 485

\bibitem{}
 Gizis J.E., Kirkpatrick J.D., Burgasser A., Reid I.N., Monet D.G., 
 Liebert J., Wilson J.C., 2001, ApJ, 551, L163
 
\bibitem{}
 Goldreich P., Tremaine S., 1980, ApJ, 241, 425 
 
\bibitem{}
 Gould A., Rix H-W., 2000, ApJ, 532, L29
 
\bibitem{}
 Haisch K.E., Lada E.A., Lada C.J., 2001, ApJ, 553, L153
 
\bibitem{}
 Halbwachs J.L., Arenou F., Mayor M., Udry S., Queloz D., 2000, A\&A, 355, 581 
 
\bibitem{} 
 Hartmann L., Calvet N., Gullbring E., D'Alessio P., 1998, ApJ, 495, 385
 
\bibitem{}
 Johnstone D., Hollenbach D., Bally J., 1998, ApJ, 499, 758
 
\bibitem{}
 K\"onigl A., 1991, ApJ, 370, L39 
  
\bibitem{}
 Kirkpatrick J.D. et al., 1999, ApJ, 519, 802

\bibitem{}
 Kirkpatrick J.D. et al., 2000, AJ, 120, 447

\bibitem{}
 Lin D.N.C., Bodenheimer P., Richardson D.C., 1996, Nature, 380, 606
 
\bibitem{}
 Lin D.N.C., Papaloizou J.C.B., 1986, ApJ, 309, 846 
 
\bibitem{}
 Lin D.N.C., Pringle J.E., 1990, ApJ, 358, 515 
 
\bibitem{} 
 Marcy G.W. et al., 2001, ApJ, 555, 418 
 
\bibitem{}
 Marcy G.W., Butler R.P., 2000, PASP, 112, 137  
 
\bibitem{}
 Mart\'\i n E.L. et al., 2000, ApJ, 543, 299
 
\bibitem{}
 Osterloh M., Beckwith S.V.W., 1995, ApJ, 439, 288 
 
\bibitem{}
 Papaloizou J.C.B., Nelson R.P., Masset F., 2001, A\&A, 366, 263  
 
\bibitem{}
 Reipurth B., Clarke C.J., 2001, AJ, 122, 432
 
\bibitem{}
 Reipurth B., Clarke C.J., Delgado-Donate E., 2001, to appear in 
 proceedings, 12th Cambridge Workshop, Cool Stars, Stellar Systems, and the Sun 
 
\bibitem{}
 Shu F.H., Johnstone D., Hollenbach D., 1993, Icarus, 106, 92  
 
\bibitem{}
 Stassun K.G., Mathieu R.D., Mazeh T., Vrba F.J., 1999, AJ, 117, 2941 
 
\bibitem{}
 Syer D., Clarke C.J., 1995, MNRAS, 277, 758
 
\bibitem{}
 Taam R.E., Spruit H.C., 2001, ApJ, 561, 329
 
\bibitem{} 
 Trilling D.E., Benz W., Guillot T., Lunine J.I., Hubbard W.B.,
 Burrows A., 1998, ApJ, 500, 428

\bibitem{}
 Trilling D., Lunine J.I., Benz W., 2001, A\&A, submitted   

\end{thebibliography}
\end{document}